# The effect of $O_2$ impurities on the low temperature radial thermal expansion of bundles of closed single-walled carbon nanotubes.

A.V. Dolbin[1], V.B. Esel'son[1], V.G. Gavrilko[1], V.G. Manzhelii[1], S.N. Popov[1], N.A.Vinnikov[1], B. Sundqvist[2].

[1] Institute for Low Temperature Physics & Engineering NASU, Kharkov 61103, Ukraine
[2] Department of Physics, Umea University, SE - 901 87 Umea, Sweden

**Abstract.**

The effect of oxygen impurities upon the radial thermal expansion $\alpha_r$ of bundles of closed single-walled carbon nanotubes has been investigated in the temperature interval 2.2-48 K by the dilatometric method. Saturation of bundles of nanotubes with oxygen caused an increase in the positive $\alpha_r$-values in the whole interval of temperatures used. Also, several peaks appeared in the temperature dependence $\alpha_r(T)$ above 20 K. The low temperature desorption of oxygen from powders consisting of bundles of single-walled nanotubes with open and closed ends has been investigated.

## 1. Introduction

Carbon nanotubes (CNT) rank among the most promising objects of fundamental and applied research. Owing to their unique structures and extraordinary mechanical, electric and thermal properties, CNTs hold a considerable potential for extensive applicability in various fields of human activity – from high-speed nano-dimensional electronics and biosensors to hydrogen power engineering and developments for ecological purposes. It is known that doping of carbon nanomaterials (fullerites [1] and nanotubes [2]) with impurities, including gaseous ones, has a considerable effect on their properties and hence on the characteristics of products and devices based on these materials. The penetration of $O_2$ molecules into bundles of single-walled nanotubes (SWNTs) affects drastically the properties of these SWNT systems, changing, for example, their conductivity by several orders of magnitude [2]. However, the influence of the $O_2$ impurity on the thermal properties of SWNT bundles, in particular their thermal expansion, still remains obscure.

It has been shown [3-6] that doping a system consisting of SWNT bundles with gases causes sharp changes in both the magnitudes and the sign of its radial thermal expansion $\alpha_r(T)$. This is due to the joint effect of several factors. Firstly, the impurity molecules sitting at the surface and inside the CNTs suppress the lowest-frequency transverse vibrations of the quasi-two-dimensional carbon walls of the nanotubes. These vibrations are characterized by negative Grüneisen coefficients [7], which determines their dominant negative contribution to the thermal expansion at low temperatures. The suppression of the transverse vibrations by gas impurity molecules reduces the negative contribution and increases the radial thermal expansion of the SWNT bundles. Another factor affecting the thermal expansion of gas-doped SWNT bundles is connected with temperature variations that provoke a spatial redistribution of the gas impurity molecules localized in different areas of the SWNT bundles and having different energies of binding to the CNTs. This shows up as peaks in the temperature dependence of $\alpha_r$. The saturation of SWNT bundles with He impurities increases the negative values of $\alpha_r(T)$ below 3.7 K, which is due to the tunneling character of the positional rearrangement of the He atoms [6].

In this study the radial thermal expansion of $O_2$-saturated bundles of single-walled carbon nanotubes with closed ends (c-SWNTs) was investigated in the interval $T$=2.2-48K by the dilatometric method. To interpret the results obtained, we needed some information about the concentration and the spatial arrangement of the $O_2$ molecules in the SWNT bundles. Such information was obtained by investigating the temperature dependence of $O_2$ desorption from bundles of closed and open SWNTs saturated with oxygen.

## 2. Low temperature desorption of oxygen impurities from carbon nanotubes.

The $O_2$ desorption from the SWNT powder was investigated in the temperature interval 50-133 K using a special cryogenic device whose design is described elsewhere [3] together with the measuring technique used. Two samples were used – the starting SWNT powder (CCVD method, Cheap Tubes, USA) and SWNT powder after an oxidative treatment was applied to open the ends of the nanotubes. The oxidative treatment is detailed in [3]. It should be noted that the oxidative-treated sample was used only to investigate desorption. The samples of c-SWNT and o-SWNTs were saturated with oxygen by the same procedure. The used $O_2$ gas was 99.98% pure and contained ≤0.02% $N_2$ as an impurity. The starting masses of the c-SWNT and o-SWNT samples were 41.6 mg and 67.4 mg, respectively. Prior to measurement, each sample was evacuated for 72 hours directly in the measuring cell of the device to remove possible gas impurities. Then the cell with the sample was filled with oxygen at room temperature to the pressure 23 Torr and cooled slowly (for 10 hours) down to 46 K. In the process of cooling the $O_2$ gas was fed to the cell in small portions as soon as the previous portion was absorbed by the SWNTs. Thus, the pressure in the cell remained no higher than the equilibrium pressure of $O_2$ vapor at each temperature. This saturation procedure allowed the maximum possible filling of all saturation-accessible positions in the SWNT bundles and on the other hand it prohibited condensation of $O_2$ vapor on the cell walls. At $T$=46 K the equilibrium pressure in the cell with the sample was 0.01 Torr, which was considerably lower than the equilibrium pressure of $O_2$ vapor at this temperature (0.04 Torr [8]). After this, the $O_2$ desorption from the nanotubes was investigated. The quantities of desorbed gas were measured during stepwise heating of the SWNT powder. The oxygen released on heating was taken to an evacuated calibrated vessel whose internal pressure was measured using a capacitive MKS-627B pressure transducer. The gas was withdrawn at each temperature of the sample until the gas pressure over the sample decreased to 0.01 Torr. Then the measurement procedure was repeated at the next temperature point.

A diagram of the desorbed $O_2$ quantities (mole per mole of SWNT powder, i.e. the number of $O_2$ molecules per carbon atom) is shown in Fig.1.

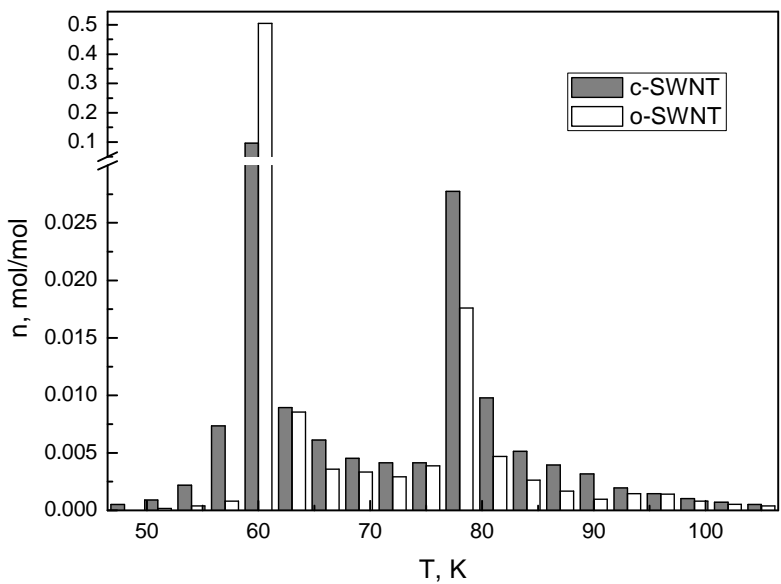

Fig.1. The relative (mol/mol) $O_2$ quantity desorbed from c-SWNTs (solid columns) and o-SWNTs (empty columns) saturated with oxygen.

It should be noted that the quantities of $O_2$ desorbed from the SWNT samples were equal, within the experimental error, to the quantities of $O_2$ sorbed by the samples on saturation, which points to a practically complete removal of the $O_2$ impurity from the sample. The reversibility of the sorption is conclusive evidence for its non-chemical origin because complete desorption of oxygen at temperatures below 110 K can only occur with physical sorption.

It was found previously [3,5] that the air-oxidative treatment of a powder of SWNT bundles led to opening the CNT ends and thus enhanced the sorptive capacity of the bundles for Xe atoms [3] and $N_2$ molecules [5]. A similar effect of such treatment might be expected for $O_2$ sorption as well. Indeed, the investigation of the sorptive capacity of the oxidized SWNTs showed that the absorbed quantity of $O_2$ increased almost threefold (see Fig.1). The quantity of $O_2$ desorbed from the treated SWNT powder increased sharply in the temperature interval 57-63K. A similar growth of the low temperature maximum in the desorption diagram was also observed for the $N_2$- saturated o-SWNT sample [5]. This is most likely because, firstly, the oxidative treatment separates nanotubes in the bundle [9] and thus increases the sorption-accessible area at the outer surface of the SWNT bundles where the energy of binding to the impurity molecules is lower in comparison with the grooves at the CNT surface [10]. Secondly, after the oxidative treatment the $O_2$ molecules are able to penetrate into the internal cavities of the CNTs through their open ends or through holes formed in the cylinder walls. It is shown theoretically [10] that the $O_2$ molecules that are located in the internal cavity along the nanotube axis and do not contact the CNT walls have much lower binding energies than the molecules localized near the inner walls. The binding energy of the $O_2$ molecules localized inside the CNTs and having no contact with the internal surface is comparable to that of the molecules forming the first layer at the outer surface of the bundles, which accounts for the highest peak of $O_2$ desorption from the o-SWNT bundles at low temperatures (57-63K).

The total quantities of $O_2$ desorbed from the starting $O_2$-saturated SWNT powder and from the oxidative-treated SWNT powder are given in Table 1.

Table 1. The total quantities of gases desorbed from c-SWNTs and o-SWNTs (mole per mole and mass.%).

| Impurity | c-SWNT | | o-SWNT | |
|---|---|---|---|---|
| | mol/mol, % | mass % | mol/mol, % | mass % |
| $H_2$ [4] | 10.0 | 1.67 | 8.07 | 1.35 |
| Xe [3] | 1.64 | 7.38 | 4.71 | 21.2 |
| $N_2$ [5] | 11.2 | 26.1 | 46.4 | 108 |
| $O_2$ | 18.1 | 42.27 | 56 | 149.4 |

The somewhat higher concentration of the sorbed $O_2$ impurity can be attributed to the higher energies of the $O_2$-$O_2$ interaction (9.2 kJ/mol [8]) in comparison with the $N_2$-$N_2$ one (6.8 kJ/mol [11]). This difference gives the $O_2$ molecules more chances to form a second and subsequent layers at the bundle surface in comparison with $N_2$ molecules.

## 3. Radial thermal expansion of oxygen-saturated bundles of closed single-walled carbon nanotubes.

The radial thermal expansion of $O_2$-saturated closed single-walled carbon nanotubes was investigated using a low-temperature capacitance dilatometer with a 0.02 nm sensitivity [12]. The sample was prepared by layer-by-layer compressing [13] a SWNT powder (Cheap Tubes,USA) at the pressure 1.1 GPa. The technique used aligned the CNT axes in the plane perpendicular to the sample axis, which was attested by an X-ray investigation. The preparation technique is detailed in [14]. The sample was a cylinder ~7.2 mm high and ~10 mm in diameter. Prior to measurement, the cell with the sample of pressure-oriented CNTs was evacuated at room temperature for 72 hours. Then the CNTs were doped with oxygen using the procedure described in Section 1. When the saturation process was completed, the measuring cell was cooled to liquid helium temperature. The thermal expansion was measured in vacuum down to $1 \cdot 10^{-5}$ Torr.

The obtained temperature dependence of the radial thermal expansion coefficient $\alpha_r(T)$ of the c-SWNT bundles saturated with oxygen is shown in Figs. 2 a, b (curve 1).

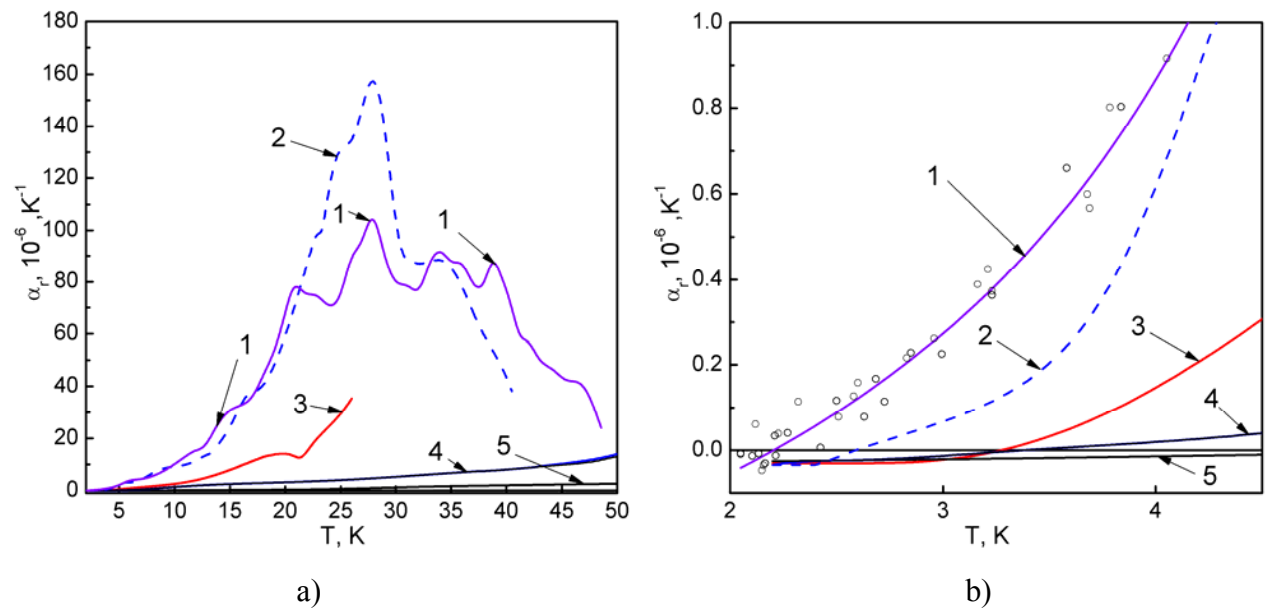

a)                                        b)

Fig. 2. Coefficient of radial thermal expansion of bundles of closed nanotubes: 1 – saturated with oxygen; 2 – saturated with nitrogen [5]; 3 – saturated with hydrogen [4]; 4 – saturated with xenon [3]; 5 – pure nanotubes [14]; a) in the temperature interval of 2,2-48 K; b) in the temperature interval of 2,2-4.5 K.

It is interesting that above 20 K the radial thermal expansion of the oxygen-saturated nanotubes (curve 1 in Fig. 2a) shows several well defined maxima. It was assumed in [3-6] that the peaks observed in the $\alpha_r(T)$ of gas-saturated SWNT bundles were caused by the positional redistribution of the gas impurity molecules at the surface of and inside SWNT bundles due to a change in the temperature. Some impurity molecules can change their energies as they move from one position to another at the surface and inside the SWNT bundles and the peaks in the temperature dependence $\alpha_r(T)$ account for such rearrangements of the impurity molecules. According to [10], the $O_2$ molecules localized in the grooves between the neighboring tubes in the c-SWNT bundles have the highest energy of binding to the bundle surface. The $O_2$ molecules forming a two-dimensional phase (layer) at the lateral surface of SWNT bundles have somewhat lower energy. The binding energies of the $O_2$ molecules forming the subsequent (second, third and so on) layers are even lower. On heating, the impurity molecules having the lowest energies of binding to the CNT surface, i.e. the molecules of the two-dimensional layers (the first and the subsequent ones), at the bundle surface are excited first. The excited molecules move from the first layer to the next ones having much lower energies of binding to the surface of SWNT bundles. This process increases the energy of the total system (SWNT bundles plus impurity molecules), which causes a peak in $\alpha_r(T)$. Further heating excites the gas molecules in the grooves at the lateral surface of SWNT bundles. They escape from the grooves and form a two-dimensional layer at the lateral surface of the bundles. Their potential energy grows, which shows up as peaks in the temperature dependence $\alpha_r(T)$. Some of the $O_2$ molecules inside the SWNT bundles penetrate into the relatively limited number of nanotubes with open ends available in the starting powder and the $O_2$ molecules can also move into the comparatively wide channels formed inside SWNT bundles between nanotubes having different diameters. Such temperature-triggered positional redistribution of the impurity molecules inside SWNT bundles can also induce peaks in the dependence $\alpha_r(T)$. The nonuniform solution of gases in the SWNT bundles leads to a considerable smearing and overlapping of the peaks in the dependence $\alpha_r(T)$. As a result, the peaks that may appear in the dependence are not all evident, which makes a detailed interpretation of the results rather difficult.

Of the gases (He, $H_2$, $N_2$, Xe) investigated previously [3-6], the effect of $O_2$ upon the thermal expansion is most closely similar to that of $N_2$. The peaks in the dependence $\alpha_r(T)$ of the $N_2$-SWNT and $O_2$-SWNT systems appear in very similar temperature intervals (Fig.2), though in the case of $N_2$ the interval is somewhat narrower. Because of the narrower temperature region of the $\alpha_r(T)$ peaks in the $N_2$-SWNT system, more peaks may overlap and therefore be partly unobservable (Fig. 2). We should also note that the saturation of SWNT bundles with oxygen led to a practically complete disappearance of the region of negative $\alpha_r(T)$-values. This is most likely because the $O_2$ concentration is relatively higher in the $O_2$-cSWNT system (Table 1) than in the other gases, and the interaction between the $O_2$ molecules and the CNT surface is appreciably stronger [10]. Together these factors suppress effectively the low temperature transverse vibrations of the quasi-two-dimensional CNT walls which are responsible for negative $\alpha_r(T)$.

It is obvious that the contribution of the positional redistribution of impurity molecules to the thermal expansion coefficient $\alpha_r(T)$ is not proportional to the impurity concentration. Variations of the $O_2$ concentration in the $O_2$-cSWNT solution cause nonuniform changes in the positions of the $O_2$ molecules in the SWNT bundles, which correspondingly affects the temperature dependence of the contribution made to $\alpha_r(T)$ by the positional $O_2$ redistribution. To decrease partially the $O_2$ concentration, the sample was heated to T=63 K, which let us remove mainly the $O_2$ molecules weakly bound to the CNTs (see Fig.1). The sample was held at T=63 K until the $O_2$ desorption proceeding at this temperature was completed (i.e., until the pressure in the measuring cell became $1\times10^{-5}$ Torr). The sample was then cooled to T=2.2 K and the thermal expansion was measured again (see Fig. 3, curve 2).

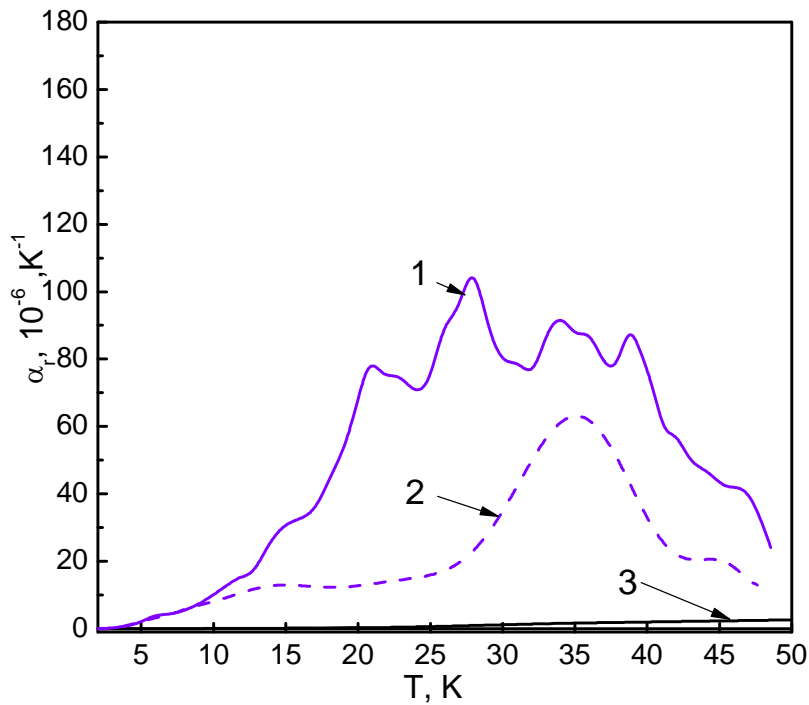

Fig. 3. The radial thermal expansion coefficient of c-SWNT bundles: 1 – saturated with $O_2$, 2 – after a partial removal of oxygen at T=63 K, 3 – pure CNTs.

It is seen that the partial removal of the oxygen impurity at $T$=63 K resulted mainly in a considerable suppression of the contribution of the $O_2$ molecules whose positional rearrangement caused a comparatively small change in the energy of the system. When the $O_2$ molecules were removed from these positions, the low temperature peaks of $\alpha_r(T)$ disappeared (at $T$=21 K and 28 K). As was expected, the partial desaturation had considerably less influence on the high-temperature peaks of $\alpha_r(T)$. For example, at $T$=35 K and 40 K these peaks transformed into a single lower peak. The effect of the partial removal of the $O_2$ impurity upon the temperature dependence $\alpha_r(T)$ of the sample was even weaker outside the peak region (below $T$=9 K). This may indicate that the thermal expansion of the SWNT sample outside the peak region is influenced mainly by the $O_2$ molecules that are localized at the sites with high binding energies, i.e. in the first layer at the bundle surface, in the internal voids of the CNTs and in the grooves between the CNTs at the bundle surface.

**Conclusions**.

The temperature dependence of the radial thermal expansion coefficient $\alpha_r(T)$ of closed single-walled carbon nanotubes saturated with oxygen has been measured in the temperature interval T=2.2-48K using the dilatometric method. Saturation of SWNT bundles with oxygen led to a sharp increase in the magnitudes of $\alpha_r$ in the whole range of temperatures investigated. The reason may be that the $O_2$ molecules decrease the negative contribution to the thermal expansion made by the transverse acoustic CNT vibrations perpendicular to the nanotube surface. The more appreciable suppression of the negative contribution in comparison with other gas impurities is attributed to the relatively high $O_2$ concentration (18 mol. %) in the SWNT bundles, as well as to the rather strong interaction between the $O_2$ molecules and the CNTs.

The temperature dependence $\alpha_r(T)$ of the $O_2$-saturated SWNT bundles has several peaks in the interval T=20-45K. They may be caused by a positional redistribution of the $O_2$ molecules at the SWNT surface and inside some nanotubes. As the concentration of oxygen was reduced through its partial desorption at T=63K, the values of $\alpha_r(T)$ decreased at $T$>9K. The analysis of the $O_2$ desorption effect on the thermal expansion of the $O_2$-cSWNT solution shows that in the interval $T$=15-45K the magnitude and the dependence of $\alpha_r(T)$ are mainly determined by the positional rearrangement of the $O_2$ molecules whose interaction with the CNTs is relatively weak. It is likely that below 9K the thermal expansion is mainly contributed by the $O_2$ molecules strongly bound to the CNTs, i.e. the $O_2$ molecules localized in the grooves of the bundles and the $O_2$ molecules forming the first layer at the bundle surface and on the inner CNT walls.

The effects produced by the sorbed oxygen and nitrogen upon the radial thermal expansion of SWNT bundles have been compared qualitatively.
The desorption of oxygen from a powder of SWNTs with open and closed ends has been investigated in the temperature interval 50-133 K.

The air-oxidative treatment of SWNT bundles aimed at opening the CNT ends 3.1 times enhanced the sorptive capacity of the sample for oxygen in comparison with the starting SWNT powder.

The authors are indebted to the Science & Technology Center of Ukraine (STCU) for the financial support of the study (Project # 5212).

**Bibliography**

1. E.A. Katz, D. Faiman, S. Shtutina, N. Froumin, M. Polak, A.P. Isakina, K.A. Yagotintsev, M.A. Strzhemechny, Y.M. Strzhemechny, V.V. Zaitsev, S.A. Schwarz, *Physica B* **304**, 348 (2001).

2. P.G. Collins, K. Bradley, M. Ishigami, A. Zettl, *Science* **287,** 1801 (2000).

3. A.V. Dolbin, V.B. Esel'son, V.G. Gavrilko, V.G. Manzhelii, N.A. Vinnikov, S.N. Popov, N.I. Danilenko and B. Sundqvist, *Fiz. Nizk. Temp.* **35**, 613 (2009) *[Low Temp. Phys. 35, 484 (2009)]*.

4. A.V. Dolbin, V.B. Esel'son, V.G. Gavrilko, V.G. Manzhelii, S.N. Popov, N.A. Vinnikov and B. Sundqvist, *Fiz. Nizk. Temp.* **35**, 1209 (2009) *[Low Temp. Phys. 35, 939 (2009)]*.

5. A.V. Dolbin, V.B. Esel'son, V.G. Gavrilko, V.G. Manzhelii, S. N. Popov, N.A. Vinnikov and B. Sundqvist, *Fiz. Nizk. Temp.* **36**, 465 (2010) *[Low Temp. Phys. 36, 365 (2010)]*.

6. A.V. Dolbin, V.B. Esel'son, V.G. Gavrilko, V.G. Manzhelii, S.N. Popov, N.A. Vinnikov and B. Sundqvist, *Fiz. Nizk. Temp.* **36**, 797 (2010).

7. P.K. Schelling and P. Keblinski, *Phys. Rev. B* **68**, 035425 (2003).

8. S. Ayoma, E. Kanda, *J. Chem. Soc. Jap.* **55**, p. 23 (1934).

9. M.T. Martínez , M.A. Callejas, A.M. Benito, M. Cochet, T. Seeger, A. Ansón, J. Schreiber, C. Gordon, C. Marhic, O. Chauvet, J.L.G. Fierro, W.K. Maser, *Carbon* **41** 2247 (2003).

10. H. Ulbricht, G. Moos, and T. Hertel, *Phys. Rev. B* **66**, 075404 (2002).

11. G.T. Furukava, and R.E. McCoskey, NACA Tech. Bote, No. 2969, p.30 (1953).

12. A.N. Aleksandrovskii, V.B. Esel`son, V.G. Manzhelii, B.G. Udovichenko, A.V. Soldatov and B. Sundqvist, *Fiz. Nizk. Temp.* **23**, 1256 (1997) *[Low Temp. Phys. 23, 943 (1997)]*.

13. N. Bendiab, R. Almairac, J. Sauvajol and S. Rols, *J. of Appl. Phys.* **93**, 1769 (2002).

14. A.V. Dolbin, V.B. Esel'son, V.G. Gavrilko, V.G. Manzhelii, N.A. Vinnikov, S.N. Popov and B. Sundqvist, *Fiz. Nizk. Temp.* **34**, 860 (2008) *[Low Temp. Phys. 34, 678 (2008)]*.